\documentclass[reprint,
 amsmath,amssymb,
 aps,
]{revtex4-2}

\usepackage{graphicx}
\usepackage{dcolumn}
\usepackage{bm}
\usepackage{mathtools}
\mathtoolsset{showonlyrefs,showmanualtags}
\usepackage{color}
\usepackage{soul}
\newcommand{\reserve}[1]{}

\newcommand{\dif}{\mathrm{d}}
\newcommand{\ecoli}{{\it E. coli} }
\newcommand{\opt}{\mathrm{OPT}}
\newcommand{\Fopt}{F_{\opt}}

\newcommand{\Fexp}{F_{\mathrm{EXP}}}
\newcommand{\sref}{\mathrm{ref}}

\newcommand{\Prob}{\mathbb{P}}
\newcommand{\Expect}{\mathbb{E}}
\newcommand{\Transpose}{\mathbb{T}}

\newcommand{\eqnref}[1]{Eq. \eqref{#1}}

\newcommand{\pos}{\xi}
\newcommand{\dir}{X}
\newcommand{\obs}{Y}
\newcommand{\post}{Z}
\newcommand{\pst}{\bar{p}}
\newcommand{\W}{W}
\newcommand{\gain}{K}

\begin{document}

\preprint{APS/123-QED}

\title{A connection between bacterial chemotactic network and optimal filtering}

\author{Kento Nakamura}
\author{Tetsuya J. Kobayashi}%
\altaffiliation[Also at ]{Institute of Industrial Science, the University of Tokyo}
 \email{tetsuya@mail.crmind.net}
\affiliation{%
 Department of Mathematical Informatics, Graduate School of Information Science and Technology, the University of Tokyo
}%




\date{\today}

\begin{abstract}
The chemotactic network of {\it Escherichia coli} has been studied extensively both biophysically and information-theoretically.
Nevertheless, the connection between these two aspects is still elusive. 
In this work, we report such a connection by showing that a standard biochemical model of the chemotactic network is mathematically equivalent to an information-theoretically optimal filtering dynamics.
Moreover, we demonstrate that an experimentally observed nonlinear response relation can be reproduced from the optimal dynamics.
These results suggest that the biochemical network of \ecoli chemotaxis is designed to optimally extract gradient information in a noisy condition.
\end{abstract}

\maketitle

Living things have developed sensory systems to behave appropriately in changing environments.
One of the most-analyzed such systems is the sensory system of {\it Escherichia coli} for chemotaxis.
In \ecoli chemotaxis, a cell obtains information of a spatial  gradient of a ligand from the temporal change in the ligand concentration that it experiences by swimming in the gradient.
An \ecoli cell can sense a positive change in the ligand concentration when it swims along the direction of the gradient and vice versa.
The swimming trajectory of \ecoli consists of a series of ballistic swimming called run interrupted with random reorientations of direction called tumbling.
By inhibiting the frequency of tumbling when it senses a positive change in an attractant concentration, the \ecoli cell can elongate the run length toward the direction of the higher concentration.

The mechanism of the sensory system has been intensively studied both experimentally and theoretically.
Experimental studies have revealed the response of \ecoli to various temporal profiles of concentration by measuring behaviors of motor rotation \cite{block1982impulse,block1983adaptation} and signaling molecules \cite{sourjik2002receptor,sourjik2004functional}.
Theoretical studies have proposed and analysed biochemical models that can reproduce properties of the experimentally observed responses such as high sensitivity to weak changes in concentration \cite{bray1998receptor,sourjik2004functional,keymer2006chemosensing,mello2007effects} and sensory adaptation \cite{barkai1997robustness}.
Based on these works, Tu \textit{et al} proposed a simplified biochemical model \cite{tu2008modeling}, which can explain various aspects of the responses simultaneously \cite{tu2013quantitative}.
This biochemical model has been widely employed for various purposes such as analysis of sensory-motor coordination \cite{jiang2010quantitative}, fold-change detection \cite{shoval2010fold,olsman2016allosteric}, and thermodynamics of sensory adaptation \cite{lan2012energy}.

In Tu's model \cite{tu2008modeling}, 
the sensory system consists of receptor complexes, each of which takes either active or inactive state.
Active receptors send a signal via mediator proteins and control the rotation of flagellar motors.
The ratio of active receptors, termed receptor activity $a_{t}$, is subjected to a feedback regulation through receptor modification characterised by methylation level $m_{t}$.
The receptor activity $a_{t}$ is determined by the free energy difference $f_{t}$ between active and inactive states:
\begin{equation}
a_{t} = \frac{1}{1+\exp(f_{t})}.    \label{eq:tu_activity}
\end{equation}
The free energy difference $f_{t}$ comprises the additive effect of the methylation level $m_{t}$ and of the ligand concentration $[L]_{t}$ as
\begin{align}
    f_{t} = N(-\alpha m_{t} + \log [L]_{t}),\label{eq:tu_free-energy}
\end{align}
where we omit a constant term and $N, \alpha > 0$ are biochemical constants.
Equations \eqref{eq:tu_activity} and \eqref{eq:tu_free-energy} take the form of the Monod-Wyman-Changeux (MWC) model describing allostery \cite{monod1965nature} where $N$ specifies the receptor cooperativity producing high sensitivity \cite{sourjik2004functional,keymer2006chemosensing,mello2007effects}.
The methylation level $m_{t}$ is modulated by the receptor activity $a_{t}$ as
\begin{align}
    \frac{\dif m_{t}}{\dif t} = F(a_{t}), \label{eq:tu_methylation}
\end{align}
where $F$ is assumed to be a monotonically decreasing function.
Since $\dif a_{t}/\dif m_{t} > 0$ and $F'(a_{t})<0$, the dynamics of the methylation level $m_{t}$ with the function $F$ constitutes a negative feedback regulation over the receptor activity $a_{t}$.
Due to the negative feedback, this biochemical network displays the sensory adaptation \cite{barkai1997robustness}, that is, when the concentration $[L]_{t}$ is stationary, the receptor activity converges to a single value $\bar{a}$ such that $F(\bar{a})=0$ which is independent of background concentration.

Although the biochemical model captures the integral parts of the sensory system and its behaviors, there is room for discussion from the view point of noise tolerance.
Because the sensory system relies on stochastic ligand-receptor interactions and receptor modifications, sensing signal inevitably contains noise.
This noise would cause a fatal influence on the chemotactic performance because it can bury the actual temporal changes in concentration and could end up with misdirections of the motor control.
Therefore, the sensory system of \ecoli is expected to have a certain noise filtering property, and several works have investigated impacts of noise in information transmission and favorable traits for noise filtering \cite{andrews2006optimal}.
However, these works focused on linear response by ignoring the underlying biochemical network and resultant nonlinear properties of the \ecoli sensory system.
Even though some others considered a possible biochemical implementation of an ideal noise-immune system based on nonlinear filtering theory \cite{kobayashi2010implementation}, the correspondence with actual biological systems, especially that of the gradient sensing in chemotaxis, is still elusive.

In this paper, we utilize nonlinear filtering theory to derive noise tolerant gradient sensing dynamics and demonstrate its biochemical implementation in \ecoli's cell.
In particular, we find that the derived ideal noise-filtering system excellently coincides with Tu's biochemical model for the \ecoli sensory system \cite{tu2008modeling} and reproduces a nonlinear response relation measured experimentally.

As a minimal model of the temporal gradient sensing, we consider a run-tumble motion of \ecoli on one dimensional axis along with monotonically increasing ligand concentration.
This assumption is mainly due to the limited capacity of the cell that may not be able to recognize the three dimensional physical space.
Let $\pos_{t} \in \mathbb{R}$ and $\dir_{t} \in \{-1,+1\}$ be the location and the direction of swimming at time $t\in [0,\infty)$.
We assume that an \ecoli cell runs ballistically with a constant speed $v > 0$ as $\dif \pos_{t}/\dif t = v \dir_{t}$ and that each run and its direction is interrupted by a stochastic tumbling motion. 
By approximating the tumbling motion by an instantaneous event\footnote{we assume that the change of direction occurs only at the tumbling event, in other words, we neglect directional change in run phase caused by viscosity},
we model the random changes in direction $\dir_{t}$ due to tumbling with a continuous-time Markov chain:
\begin{align}
    \frac{\dif \bm{p}_{t}}{\dif t} = \left(\begin{array}{cc}
        -r^- & r^+ \\
        r^- & -r^+
    \end{array}\right)\bm{p}_{t} \label{eq:state},
\end{align}
 where $\bm{p}_{t}=(\Prob(\dir_{t}=+1),\Prob(\dir_{t}=-1))^{\Transpose}$, and $r_+$ and $r_-$ are the time-independent transition rates from $-1$ to $+1$ and from $+1$ to $-1$, respectively.
Note that the transition rate of direction $\dir_{t}$ would be smaller than the rate of tumbling event because each tumbling does not always lead to the flipping of the direction.

Next, we assume that the ligand concentration depends exponentially on the location as $[L]_{t} \propto \exp(c\pos_{t})$
where $c>0$ is a constant. 
This assumption is natural because the spatial distribution of a ligand typically obeys diffusion.
Then, we define a noisy sensing of the ligand by adding a noise term to the ligand-dependent term in \eqnref{eq:tu_free-energy} as 
\begin{align}
    \obs_{t}  = - \log [L]_{t} - \sqrt\sigma \W_{t} \label{eq:observation}
\end{align}
where $\W_{t}$ is the standard Wiener process and $\sigma$ is the intensity of noise.
It should be noted that $\W_{t}$ can also be interpreted as the noise from methylation \cite{korobkova2004molecular} because the methylation level $m_{t}$ additively appears in \eqnref{eq:tu_free-energy}.

\reserve{(Reserved for getting back to old representation) By applying the nonlinear filtering theory under the above settings and assumptions \cite{jazwinski2007stochastic}, we can derive a stochastic differential equation for the posterior probability $\post_{t} := \Prob(\dir_{t}=-1\mid \obs_{0:t})$ of the direction given the time series of the noisy sensing $\obs_{0:t}$:
\begin{align}
    \frac{\mathrm{d}\post_{t}}{\mathrm{d}t}&=-R(\post_{t}-\pst)+\gain\post_{t}(1-\post_{t})\circ\frac{\mathrm{d}\obs_{t}}{\mathrm{d}t}, 
\end{align}
where $R:= r^++r^-$, $\pst= r^-/(r^++r^-)$, $\lambda := 2vc$, and $\circ$ is the Stratonovich integral (See supplementary material for details of derivation).}
By applying the nonlinear filtering theory under the above settings and assumptions \cite{jazwinski2007stochastic}, we can derive the following stochastic differential equation as 
\begin{align}
    \frac{\mathrm{d}\post_{t}}{\mathrm{d}t}&=-R(\post_{t}-\pst)+\gain\post_{t}(1-\post_{t})\circ\frac{\mathrm{d}\obs_{t}}{\mathrm{d}t}, \label{eq:filter_expectation}
\end{align}
where $\circ$ is the Stratonovich integral (See supplementary material (SM) for details of derivation).
This equation describes the posterior probability $\post_{t} = \Prob(\dir_{t}=-1\mid \obs_{0:t})$ of the descending direction given the time series of the noisy sensing $\obs_{0:t}:=\{\obs_{t'}|t'\in [0,t]\}$ when its parameter values matches those of tumbling, run, gradient, and noise as $R= R_{\opt}:=r^++r^-$, $\pst=\pst_{\opt}:= r^-/(r^++r^-)$, $\gain = \gain_{\opt}:= 2vc/\sigma$.

Under this set of the optimal parameter values, the first term represents the prediction based on a prior knowledge about switching dynamics of direction $\dir_{t}$ (\eqnref{eq:state}).
Thereby, without the second term (sensing signal), $\post_{t}$ converges to the stationary probability of the direction $\pst$ for $t \to \infty$.
The second term corresponds to the update of the posterior by new observation (\eqnref{eq:observation}).
The optimal gain of this term, $\gain_{\opt}$, describes the signal-to-noise ratio because $\sigma$ and $2vc$ specifies the noise intensity and the steepness of the temporal change in the ligand concentration during swimming, respectively.
\reserve{The posterior $\post_{t}$ is optimal in a sense that $\Expect[\dir_{t} \mid \obs_{0:t}] = 1-2 \post_{t}$ is the minimum mean square estimator for $\dir_{t}$.}
We call the dynamics of $\post_{t}$ described by \eqnref{eq:filter_expectation} the filtering dynamics hereafter.


Next, we reveal the relation between the filtering dynamics and the biochemical network of \ecoli chemotaxis by demonstrating that \eqnref{eq:filter_expectation} can be equivalent to Eqs. \eqref{eq:tu_activity},\eqref{eq:tu_free-energy}, and \eqref{eq:tu_methylation} if noise is neglected.

To this end, we introduce a coordinate transform from the posterior probability $\post_{t}$ to the log-likelihood ratio $\theta_{t} := \log \post_{t}/(1-\post_{t})$.
From the chain rule for derivatives, $\dif \theta_{t}/\dif t = (\dif \theta_{t}/\dif Z_{t})(\dif Z_{t}/\dif t)$, we obtain the following equivalent representation of the filtering dynamics:
\begin{align}
\frac{\mathrm{d}\theta_{t}}{\mathrm{d}t}&=R\frac{\post_{t}-\pst}{\post_{t}(1-\post_{t})}-\gain\circ\frac{\mathrm{d}\obs_{t}}{\mathrm{d}t}. \label{eq:filter_natural}
\end{align}
By further defining a new variable $\mu_{t}$ for the prediction dynamics as
\begin{align}
    \frac{\dif \mu_{t}}{\dif t} := -\frac{R}{\kappa} \frac{\post_{t}-\pst}{\post_{t}(1-\post_{t})},  \label{eq:fd_prediction}
\end{align}
then we can formally integrate \eqnref{eq:filter_natural} as
\begin{align}
\theta_{t}&=- \kappa \mu_{t}+\gain\left[ \log [L]_{t} + \sqrt\sigma \W_{t}\right].
\label{eq:fd_likelihood-ratio}
\end{align}
where we use \eqnref{eq:observation} and $\kappa>0$ is an arbitrary constant. 
Finally, $\post_{t}$ in \eqnref{eq:fd_prediction} can be obtained by the  inverse transformation from $\theta_{t}$ to $\post_{t}$:
\begin{align}
    \post_{t} &= \frac{1}{1+\exp(\theta_{t})} \label{eq:fd_posterior}.
\end{align}

These transformations unveil that Eqs. \eqref{eq:fd_posterior},\eqref{eq:fd_likelihood-ratio}, and \eqref{eq:fd_prediction} for the filtering dynamics are equivalent to Eqs. \eqref{eq:tu_activity},\eqref{eq:tu_free-energy}, and \eqref{eq:tu_methylation} for the biochemical model of \ecoli chemotaxis, respectively (see also table S1 in SM for comparison).

The posterior probability $\post_{t}$ corresponds to the receptor activity $a_{t}$ and they are both described by the sigmoidal function of $\theta_{t}$ and $f_{t}$, respectively.
The log-likelihood ratio $\theta_{t}$ is determined by the logarithm of the ligand concentration $[L]_{t}$ and the prediction term $\mu_{t}$, which corresponds to the dependence of the free energy difference $f_{t}$ on the ligand concentration $[L]_{t}$ and the methylation level $m_{t}$ in \eqnref{eq:tu_free-energy}.
Finally, the dynamics of prediction term $\mu_{t}$ corresponds to that of the methylation level $m_{t}$.

\begin{figure}
\includegraphics[width=8cm]{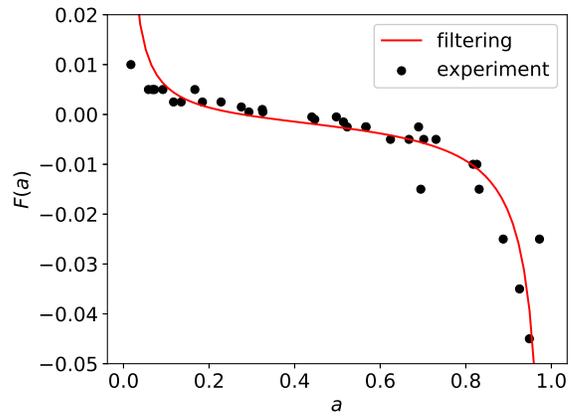}
\caption{Theoretically derived $\Fopt$ (red curve) fitted to the experimentally obtained $\Fexp$ (black points) \cite{shimizu2010modular}.
$\Fopt$ in the figure is obtained by modulating two parameters, $R$ and $\pst$, as $R/\kappa \approx 2.8 \times 10^{-3}$ and $\pst \approx 0.28$ (see also SM for the fitting procedure).
}
\label{fig:comparison}
\end{figure}

Because the right-hand-side of \eqnref{eq:fd_prediction} is a decreasing function of $\post_{t}$ in the same way as the feedback function $F(a_{t})$ of $m_{t}$, $\mu_{t}$ works as a negative feedback component to $\post_{t}$.
Even though $F(a_{t})$ in Tu's model cannot be determined biochemically but inferred only experimentally, the filtering dynamics provide a concrete functional form for the feedback function, $\Fopt(\post):=-(R/\kappa)\cdot (\post-\pst)/(\post(1-\post))$.
Thus, if \ecoli has developed the sensory system being tolerant to sensing noise near optimally, the feedback function $F$ describing the methylation dynamics can have a similar form as $\Fopt$.
To test this expectation, we compare the feedback function $\Fexp$ inferred experimentally by a FRET measurement \cite{shimizu2010modular} with the theoretically predicted $\Fopt$ by adjusting two free parameters $R/\kappa$ and $\pst$.
Figure \ref{fig:comparison} shows a notable agreement between the experimental data and  theoretical prediction.
Both $\Fexp$ and $\Fopt$ share a characteristic nonlinearity; a gentle slope around $a=0.5$ and a sharp decline near $a=1$.
This result implies that the \ecoli chemotactic network is designed structurally to be robust to the sensory noise.
In addition, because $\pst$ in $\Fopt$ corresponds to the stationary probability that the direction of swimming is down the gradient, the parameter values $\pst \approx 0.28$ obtained by fitting implies that \ecoli has a prior expectation that it likely swims up the gradient.

We further investigate whether the biochemical parameters observed experimentally in laboratory environments can satisfy the optimality in terms of filtering. 
From the fitting of $\Fopt$ to $\Fexp$, we have $R/\kappa\approx 2.8\times 10^{-3}$.
$\kappa$ can be estimated as $\kappa = \alpha N\approx 12$ by comparing \eqnref{eq:tu_free-energy} and \eqnref{eq:fd_likelihood-ratio} and by employing a previous estimate of $\alpha$ and $N$ \cite{shimizu2010modular}. 
Thus, $R$ is calculated as $R \approx 3.4\times 10^{-2}$.
In contrast, the optimal $R_{\opt}$ can be estimated from $R_{\opt}=r^++r^-$ and measurements of tumbling rate as $10^{-0.5}\leq R_{\opt}\leq 10^0 s^{-1}$ \cite{block1983adaptation,pohl2017inferring}.
Thus, the obtained biochemical parameter $R$ is much smaller than the estimate $R_{\opt}$ from tumbling measurements.

This discrepancy may be attributed to three possibilities: First, experimental conditions for the measurements of tumbling rate might not capture a wild condition where \ecoli cells are supposed to perform chemotaxis.
Recent studies suggest that swimming behavior in polymeric solutions or soft agar is different from that under a liquid condition used in most experiments \cite{patteson2015running}.
In particular, the tumbling frequency is shown to decrease with addition of polymeric molecules due to remodeling of signaling pathway downstream of sensory system or possibly due to motor load.
In such a case, $R_{\opt}$ may take smaller value.
Second, the values of $R$ might be underestimated because of the difficulty in estimating the biochemical parameter $N$.
Although we used an estimate $N=6$ in previous studies \cite{mello2007effects,tu2008modeling,shimizu2010modular}, other estimates of $N$ are larger, $N=15\sim20$ \cite{mello2007effects,endres2008variable}.
The last possibility is that the system is not or cannot be always optimized at the level of parameter values, though it is so at the level of network structure.
By considering the correspondence of $N$ with the gain $K_{\opt}$, which is determined by the speed of swimming, steepness of the gradient, and intensity of sensing noise, $N$ should not be fixed at certain value but be variable depending on environmental situations.
Several studies suggested that $N$ as well as other parameters are diversified in a population of cells for hedging environmental uncertainties \cite{sourjik2012responding}.

To perform chemotaxis under the limitation in parameter adjustment, the robustness against the mismatch of parameters could be beneficial.
We investigate whether such robustness is endowed by examining the filtering dynamics with misspecified parameter values of $K$.
We measure the performance of the dynamics using mean square error (MSE) defined as $\left[\frac{1}{T}\int_{t=0}^{T} \{\dir_{t}- (1-2Z_{t})\}^2\dif t\right]^{1/2}$ in which $1-2\post_{t}=1-2\Prob(\dir_{t}<0\mid \obs_{0:t})=\Expect[\dir_{t}\mid \obs_{0:t}]$ holds for the optimal parameter set.
We define a reference value of $K$ as $K_{\sref}:=N=6$ according to the correspondence between $K$ and $N$.
We set swimming speed to a physiologically relevant value: $v = 20 \mu \mathrm{m}\cdot\mathrm{s}^{-1}$.
The rates of directional changes are determined as $r^+ = R(1-\bar{p}), r^- = R\bar{p}$ so that the values of $R$ and $\bar{p}$ obtained by fitting in Fig. \ref{fig:comparison} become optimal.
We define the reference of the steepness of gradient as $c_{\sref}:=10^{-3}\mu\mathrm{m}^{-1}$ by taking into account conditions in previous simulation studies \cite{jiang2010quantitative}.
We also define the reference of noise intensity as $\sigma_{\sref}:=2c_{\sref}v/K_{\sref}$ such that the reference parameter $K_{\sref}$ is optimal under $c=c_{\sref}$ and $\sigma=\sigma_{\sref}$.
Note that $K_{\sref}$ is also optimal on the half-line, $(\sigma, c) = \eta (\sigma_{\sref}, c_{\sref}), \eta>0$, because $2vc/\sigma=2vc_{\sref}/\sigma_{\sref} = K_{\sref}$ holds on it.
\begin{figure}
\includegraphics[width=\linewidth]{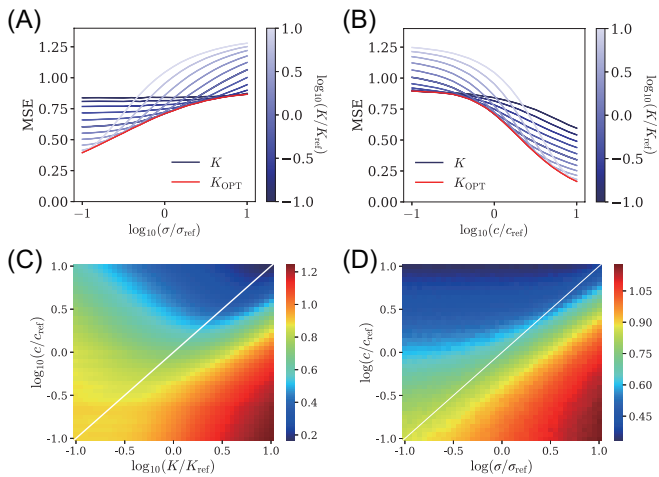}
\caption{MSE of the filtering dynamics as a function of $\sigma$ with fixed $c = c_{\sref}$ (A), as a function of $c$ with fixed $\sigma = \sigma_{\sref}$ (B),
as a function of $K$ and $c$ with fixed $\sigma=\sigma_{\sref}$ (C), and as a function of $\sigma$ and $c$ with fixed $K=K_{\sref}$ (D).
Curves in (A) and (B) represent MSEs with fixed parameter $K=K_{\sref}$ (blue) and with the optimal parameter $K=K_{\mathrm{OPT}}=2vc/\sigma$ (red).
    White lines in (C) and (D) represent the parameter region on which the parameter $K$ is set optimal i.e. $2vc/\sigma_{\sref}=K$ (C) and $2vc/\sigma=K_{\sref}$ (D).}
\label{fig:K_c_sigma}
\end{figure}


Figure \ref{fig:K_c_sigma} shows MSEs of \eqnref{eq:filter_expectation} for different $K$ as functions of $\sigma$ with fixed $c=c_{\sref}$ (A) and as functions of $c$ with fixed $\sigma=\sigma_{\sref}$ (B).
The error with fixed $K$ is always greater than or equal to that with $K$ adjusted to $K_{\opt}$.
For each fixed gain $K$, MSE monotonously increases as the signal-to-noise ratio (SNR) decreases either by the increase in the noise intensity $\sigma$ (Fig. \ref{fig:K_c_sigma}(A)) or by the decrease in gradient steepness $c$ (Fig. \ref{fig:K_c_sigma}(B)), indicating that
greater SNR than optimal one never impair the performance of the dynamics for any $K$.
We can see a similar trend in Fig. \ref{fig:K_c_sigma} (C) and (D).
These results indicate that even under the misspecification of $K$ associated with parameters $\sigma$ and $c$, the filtering dynamics still reliably and robustly estimate temporal gradient if the change in $\sigma$ and $c$ is one that increases SNR.

Small value of gain $K$ is optimized to a low SNR situation, and variation of MSE between low and high SNRs is small (Fig. \ref{fig:K_c_sigma}). 
In contrast, large $K$ adjusted to a high SNR one shows a significant variation in MSE between low and high SNR cases.
This means that low $K$ can work moderately well for most of conditions whereas large $K$ can work much better if the environmental SNR is large enough at the cost of lower performance under low SNR situations.
Thus, $K$ modulates the balance of risk-averting and -taking strategies of sensing.


The growth-dependent variability of $K$ can coordinate such risks at the level of population \cite{khursigara2011lateral}.
Moreover, $N$, which biochemically corresponds to $K$, is suggested to vary temporally at the single-cell-level \cite{endres2008variable,frank2013prolonged} via a receptor cluster rearrangement.
The integration of biochemical modeling and optimal filtering theory could work for further analysis of such a gain adaptation of cells.
This approach may also apply to other sensory systems with allosteric receptors and a negative feedback, e.g., G protein-coupled receptors for vision and EGF receptor in animal cells, whose models can be reduced to similar biochemical models to Tu's \cite{tu2008modeling}.
Furthermore, we might extend it to directly include the closed cycle between sensing of environment and the resultant actions of cells.

This research is supported by JSPS 19H05799 and JSPS 20J21362.
\bibliographystyle{apsrev4-2}

\providecommand{\noopsort}[1]{}\providecommand{\singleletter}[1]{#1}%

\end{document}